\documentclass[amssymb,twocolumn,
aps,floatfix,showpacs]{revtex4}
\usepackage{graphicx}

\begin{document}
\title{Atomic Harmonic Generation in Time-Dependent $R$-Matrix Theory}
\author{A. C. Brown}
\author{D. J. Robinson}
\author{H. W. van der Hart}
\affiliation{Centre for Theoretical Atomic, Molecular and Optical Physics,
  Queen's University Belfast, Belfast, BT7 1NN, UK.}

\date{}

\begin{abstract} We have developed the capability to determine accurate harmonic
  spectra for multielectron atoms within time-dependent $R$-matrix (TDRM)
  theory.  Harmonic spectra can be calculated using the expectation value of the
  dipole length, velocity or acceleration operator. We assess the calculation of
  the harmonic spectrum from He irradiated by 390 nm laser light with
  intensities up to $4\times 10^{14}$ Wcm$^{-2}$ using each form, including the
  influence of the multielectron basis used in the TDRM code.  The spectra are
  consistent between the different forms, although the dipole acceleration
  calculation breaks down at lower harmonics. The results obtained from TDRM
  theory are compared with results from the HELIUM code finding good
  quantitative agreement between the methods. We find that bases which include
  pseudostates give the best comparison with the HELIUM code, but models
  comprising only physical orbitals also produce accurate results.
\end{abstract}

\pacs{32.80.Rm, 31.15.A-, 42.65.Ky}
\maketitle


\section{Introduction} In recent years, harmonic generation (HG) has become one
of the richest veins of research for atomic, molecular and optical physics. Not
only has HG enabled ultrashort light pulse generation \cite{attosecond_pulse_train}, but it
has also given rise to a series of very sensitive measurements of molecular
\cite{multielectron_molecules}, atomic and even electronic dynamics \cite{multielectron_atoms}.  The
sensitive nature of HG has made it increasingly important to develop
accurate methods of modelling the process. 


Many studies aimed at describing HG  make use of the single active electron
(SAE) model \cite{sae_gavrila,sae_schafer}, 
%
a significant simplification which allows for the  efficient computation of
harmonic spectra. 
SAE methods have been used to probe the relationship between atomic structure
and HG. For instance, the Cooper minimum in Argon has been linked with the
minimum in the photoionization spectrum caused by a zero dipole moment between
the $p$ ground state wavefunction and the $d$ wavefunction of the photoionized
electron for a photon energy of around 48 eV \cite{cooper_minimum_hhg_worner,cooper_minimum_hhg_higuet}. The
minimum is observed to exist in both the photoionization spectrum and the
harmonic spectrum, and is easily described by the SAE method as it does not
depend on the interactions between different electrons. However, there have been
various studies carried out in molecular systems where multielectron dynamics
are found to be of great importance \cite{corkum_review,multielectron_molecules}. Even in
atomic systems there are features of photoionization, and hence harmonic
spectra, that are the result of electronic interactions which require a
multielectron description
\cite{brown_prl,multielectron_atoms,many_e-_hhg,multichannel_hhg}.

Over the last few years, we have developed time-dependent $R$-matrix (TDRM)
theory to model the interaction of atoms with short, intense laser pulses,
maintaining a full description of the multielectron dynamics involved
\cite{tdrm,tdrm_argon,collect_c+1}.  TDRM has recently been extended to account for
harmonic generation, and this capability was demonstrated in showing how
autoionizing resonances can affect the harmonic spectrum of Argon. The
appearance of the autoionizing resonances in these spectra is a consequence of
multielectron dynamics: the interference between the response of $3p$ and $3s$
electrons to the laser field \cite{brown_prl}.  These calculations represent an
important shift in thinking on HG: the multielectron nature of the process is
reflected in the theoretical approach, and while there are many processes that
can be adequately described using SAE methods, there are many for which this,
more rigorous, description may be required.

The determination of the harmonic spectrum can proceed through the calculation
of the time-dependent expectation value of the dipole, dipole velocity or dipole
acceleration operator. At present there is discussion about which of these
operators offers the best prediction of the harmonic response for a single atom.
Recent work \cite{dipole_gauge_madsen,velocity_hhg} has suggested that there is a natural
connection with the dipole velocity, while, commonly, the dipole acceleration
operator is used \cite{acceleration_hhg_eberly,acceleration_hhg_lappas,acceleration_hhg_burnett}, especially for the
description of high order harmonics as better resolution can be obtained for the
high energy peaks \cite{dipacc}. Much early work in the field used the dipole
length \cite{length_hhg_eberly,length_hhg_bandarage}, and up until this point the description of
HG in TDRM has been restricted to using this operator \cite{brown_prl}. We note
that the use of these various forms has been verified only within the SAE
approximation, and hence we seek herein to verify the independence of HG with
respect to the use of the dipole, its velocity or acceleration in a
multielectron system. We also assess which form offers the most numerically
stable method, particularly when used with a limited multielectron basis set.
Studies assessing the propagation of the wavefunction have demonstrated that to
obtain the most accurate results for a limited basis in the TDRM approach the
laser field is best described in the dipole length gauge \cite{tdrm_dipole_gauge}. On
the other hand, time-propagation in SAE calculations is commonly performed by
describing the laser field in the velocity gauge. This difference indicates that
we cannot necessarily rely upon knowledge gained from SAE calculations for the
assessment of TDRM calculations.

We have extended the TDRM method to calculate the harmonic spectrum from the
dipole velocity and acceleration operators simultaneously with the dipole
operator spectrum.  In this paper we cover the major theoretical aspects of this
extension, and apply the TDRM codes to He in a 390nm laser field. Helium is
chosen for three reasons. Firstly, the simple structure allows for the
systematic varying of the multielectron basis functions, the impact of which has
been assessed for TDRM in terms of photoionization \cite{tdrm_dipole_gauge}, but not for
HG. Secondly, the absence of a closed core simplifies the calculation of dipole
acceleration matrix elements, and hence we can compare spectra in all three forms. Finally, using He
allows us to benchmark our approach against a proven alternative method: we
compare our results with those obtained using the HELIUM code \cite{HELIUM}.


\section{Theory} 
\subsection{TDRM Theory}
The TDRM approach is an {\it ab initio} nonperturbative theory for describing ultrafast
atomic processes. Details of the method can be found in \cite{tdrm,tdrm_dipole_gauge},
so we only give a short overview here.
The time-dependent Schr\"{o}dinger equation for an atom containing ($N+1$)
electrons is 
\begin{equation}
\label{tdse}
i\frac{\partial}{\partial t} \Psi \left( \mathbf{X}_{N+1},t\right)= 
H\left(t\right)\Psi\left(\mathbf{X}_{N+1},t\right).
\end{equation}
The Hamiltonian, $H$, contains both the non-relativistic Hamiltonian of the
$N+1$-electron atom or ion in the absence of the laser field and the laser
interaction term.  The laser field is described using the dipole approximation
in the length form, and is assumed to be linearly polarized and spatially
homogeneous. This form provides the most reliable ionization
yields when only a limited amount of atomic structure is included
\cite{tdrm_dipole_gauge}.

We propagate a solution of the time-dependent Schr\"{o}dinger equation $\Psi$ on
a discrete time scale with time step $\Delta t$ in a Crank-Nicolson scheme. We
can write the wavefunction at a time $t_{q+1}$ in terms of the wavefunction at
the previous time step $t_q$:
\begin{equation} \label{recurs1} (H_m-E)\Psi _{t_{q+1}} = -(H_m +E)\Psi _{t_q}.
\end{equation}
Here the imaginary energy $E$ is defined as $2i/\Delta t$ and $H_m$ is the
Hamiltonian at the midpoint of the time interval, $t_{q+1/2}$. 

In $R$-matrix theory, configuration space is partitioned into an inner and outer
region. In the inner region, all electrons are within a distance
$a_{\mathrm{in}}$ of the nucleus, and full account is taken of all interactions
between all electrons. In the outer region, an ionized  electron moves beyond
the boundary $a_{\mathrm{in}}$, and thus exchange interactions between this
electron and the electrons remaining close to the nucleus can be neglected. The
ionized electron then moves in only the long-range multipole potential of the
residual $N$-electron core and the laser field.

Following \cite{tdrm_argon} we can evaluate
Eq. (\ref{recurs1}) at the boundary $a_{\mathrm{in}}$ as a
matrix equation
\begin{equation}
\label{frtmat}
\mathbf{F}\left(a_{\mathrm{in}}\right)=
\mathbf{R}(a_{\mathrm{in}})\bar{\mathbf{F}}\left(a_{\mathrm{in}}\right)+\mathbf{T}\left(a_{\mathrm{in}}\right),
\end{equation}
in which the wavefunction $\mathbf{F}$, at the boundary is described in terms of its
derivative, $\bar{\mathbf F}$, plus an inhomogeneous vector, ${\mathbf T}$,
arising from the right hand side of Eq. (\ref{recurs1}). The $R$-matrix,
$\mathbf{R}$, connects the inner and outer region wavefunction at the boundary
$a_{\mathrm{in}}$. 

Given an inner region wavefunction, $\mathbf{R}$ and $\mathbf{T}$ are evaluated
at the boundary $a_{\mathrm{in}}$. Subsequently, they are propagated outwards in
space up to a boundary, $a_{\mathrm{out}}$ where it can be assumed that the
wavefunction $\mathbf{F}$ has vanished. The wavefunction vector $\mathbf{F}$ is
set to zero and then propagated inwards to the inner region boundary. Once
$\mathbf{F}$ has been determined at each boundary point, the full wavefunction
can be extracted from the $R$-matrix equations. We can then iterate the
procedure using Eq. (\ref{recurs1}). 

\subsection{Harmonic generation} 
The electric field produced by an accelerating charge is given, using the non-relativistic Lienard-Wiechert
potentials in the far field limit, by
\begin{equation} \label{lienard} E(t)=k\left\langle\psi(t)\left|
  \frac{[p_z,H]}{i\hbar}\right|\psi(t)\right\rangle + ke  E_{\mathrm{laser}}(t),
\end{equation}
where $e$ is the electronic charge, $z$ is the laser polarization axis, $k$ is a proportionality
constant, $p_z$ is the canonical momentum and $E_{\mathrm{laser}}$ is the
electric field of the laser pulse. We can write 
\begin{equation} \label{ehren_theorem}  \left \langle \psi(t) \left| \frac{[p_z,H]}{i\hbar} \right|  \psi(t) \right \rangle =
  \frac{d}{dt} \left \langle \psi(t) | p_z | \psi(t) \right \rangle,
\end{equation}
and it follows that 
\begin{equation}
  E(t) \propto \mathbf{\ddot{d}}(t) =
  \frac{d^2}{dt^2}\langle\psi(t)|\mathbf{z}|\psi(t)\rangle.
\end{equation}
The power spectrum of the emitted radiation is then given, up to a
proportionality constant,  by
$|\mathbf{\ddot{d}}(\omega)|^2$- the Fourier transform of $\mathbf{\ddot{d}}(t)$
squared. 

Although the radiation produced is proportional to the dipole acceleration, it
is common practice in HG calculations to calculate $\mathbf{d}(\omega)$, i.e.
to use the expectation value of the dipole length instead. This is because a
simple relationship exists between $\mathbf{d}$ and $\mathbf{\ddot{d}}$ which
can be extended to include the dipole velocity form:
\begin{equation} \label{rescaling} \omega ^4|\mathbf{d}(\omega)|^2 = \omega ^2
  |\dot{\mathbf{d}}(\omega)|^2 = |\ddot{\mathbf{d}}(\omega)|^2 .  \end{equation}

Therefore the harmonic response of a single atom can be expressed in terms of the
expectation value of the dipole operator
\begin{equation}
\mathbf{d}\left(t\right)=\langle \Psi \left(t\right)
|-e\mathbf{z}|\Psi\left(t\right)\rangle,
\label{inducedip_len}
\end{equation}
or of its velocity 
\begin{equation}
  \mathbf{\dot{d}}\left(t\right)=\frac{d}{dt}\langle \Psi \left(t\right)
  |-e \mathbf{z}|\Psi\left(t\right)\rangle,
\label{inducedip_vel}
\end{equation}
or acceleration 
\begin{equation}
  \mathbf{\ddot{d}}\left(t\right)=\frac{d^2}{dt^2}\langle \Psi \left(t\right)
  |-e \mathbf{z}|\Psi\left(t\right)\rangle,
\label{inducedip_acc}
\end{equation}
where $\mathbf{z}$ is the total position operator
along the laser polarization axis.

As discussed in \cite{tdrm_dipole_gauge} the TDRM code can use either the length or
velocity gauge for the propagation of the wavefunction. While, in keeping with
the findings of \cite{tdrm_dipole_gauge}, we use the length gauge, we can still utilize
the dipole velocity matrix elements produced by the $R$-matrix suite of codes
which `seed' the TDRM code. Thus we can store both $\mathbf{z}$ and $d\mathbf{z}/dt$ and use Eqns.
(\ref{inducedip_len}) and (\ref{inducedip_vel}) directly for the determination
of the time-varying expectation values of the dipole operator and the dipole
velocity.

However, in order to calculate the expectation value of the dipole acceleration
we cannot use Eq. (\ref{inducedip_acc}) directly. Instead, using Ehrenfest's theorem,
it is possible to write the dipole acceleration
as \begin{equation} \label{ehren} \mathbf{\ddot{d}}\left ( t \right) = \langle
  \frac{\partial H} {\partial r} \rangle = \langle
  \frac{eZ\cos\theta}{\mathbf{r}\cdot\mathbf{r}} \rangle -eN_{elec}\langle \Psi
  | E(t)  | \Psi \rangle, \end{equation}
where $Z$ is the nuclear charge, $\mathbf{r}$ the total position operator,
$\theta$ the angle between $\hat{\mathbf{r}}$ and $\hat{\mathbf{z}}$ and
$N_{elec}$ the number of electrons. The second term in Eq. (\ref{ehren}) is
often seen without this factor of $N_{elec}$ as in the SAE approximation it is
just 1.  We can make a small change to the way the radial integrals are
calculated in the $R$-matrix suite which allows the calculation of $\langle
1/\mathbf{r}\cdot\mathbf{r} \rangle$ instead of $\langle \mathbf{r} \rangle$.  Then we can
use Eq. (\ref{ehren}) to calculate the dipole acceleration.  Thus, we can now
simultaneously calculate harmonic spectra using the dipole length, velocity and
acceleration operators. The propagation of the wavefunction is still carried
out in the length gauge.

The use of the acceleration form will however be restricted to He like targets.
The use of Ehrenfest's theorem (in Eqs. (\ref{ehren_theorem}) and (\ref{ehren})) requires that the
wavefunction be exact, or close to it. For general multielectron systems we
normally impose a fixed core where (at least) the first two electrons are
restricted to a single orbital.  Imposing this restriction means that the
electronic repulsion is not fully described.  More precisely, if the orbital of
electron $e_1$ is fixed, and the orbital of electron $e_2$ is not, then the
action on $e_2$ will not necessarily equal minus the reaction on $e_1$. Thus,
the commutator
\begin{equation}
  \left[(\mathbf{p_1}+\mathbf{p_2}),\frac{1}{|\mathbf{r_1}-\mathbf{r_2}|}\right]
\end{equation}
may not be guaranteed to vanish. On the other hand
\begin{equation}
  \left[(\mathbf{r_1}+\mathbf{r_2}),\frac{1}{|\mathbf{r_1}-\mathbf{r_2}|}\right]
\end{equation}
will still vanish. Thus, while the expectation value
$\langle[\mathbf{r}\cos\theta,H]\rangle$ can be calculated accurately,
$\langle\left[\left[\mathbf{r}\cos\theta,H\right],H\right]\rangle$ can not,
rendering Ehrenfest's theorem untenable. Thus, the comparisons we employ for
the simple He test case which follows can be extended to general multielectron
systems only for the dipole length and velocity forms. 


\subsection{Calculation parameters} 
\label{sec:calcparam}

The one-electron basis used for describing the residual He$^+$ in the inner
region consists of orbitals expressed in terms of $B$ splines.  The residual
He$^+$ ion is represented through a series of models of increasing complexity
\cite{tdrm_dipole_gauge}.  The basic model consists of only the
He\textsuperscript{+} $1s$ state, which we call 1T (1 {\bf T}rue state). We
also use two models comprising six states. The first is built using true
orbitals $1s$, $2s$, $2p$, $3s$, $3p$, $3d$ (6T) and the other using 5 pseudo orbitals and
the true $1s$ orbital: $1s, \overline{2}s,
\overline{2}p, \overline{3}s, \overline{3}p,
\overline{3}d$ called 6P, (6 states with {\bf P}seudostates).
Pseudostate models have been found to be more accurate in the time propagation
of the He wavefunction responding to short light fields, especially in the
velocity-gauge description of the light field.  Pseudostate models may thus
provide a better basis for the description of the ionization and HG processes,
provided that these processes are not affected by artificial resonances
introduced by the pseudostates.

The inner region radius is set at 20 a.u. which is sufficiently large to contain the
residual ion for each model we use. The outer region boundary is set at 600
a.u. to prevent any reflections of the wavefunction for the duration of the
short laser pulse employed. The set of continuum orbitals contains 80
$B$ splines for each angular momentum, $\ell$, of the continuum electron up to a
maximum value, $L_{max}=19$. Convergence testing was carried out retaining
angular momenta up to a value of $L_{max}=27$ and, while changes in the harmonic
spectra are observed, they occur at energies beyond the cutoff- outside the
region of interest here. The outer region is divided into sectors of 2 a.u.
containing 35 9th order B-splines per channel. The time step
used in the wavefunction propagation is 0.1 a.u.    

We use 390 nm laser pulses, consisting of a 3 cycle $\sin^2$ ramp-on followed by
two cycles at peak intensity, followed by a 3 cycle $\sin^2$ ramp-off (3-2-3).
We also calculate spectra for different pulse shapes and find that while the
spectra change, the comparisons between them are generally described by the
results presented below for the 3-2-3 pulse. There is one important exception to
this general observation, which is discussed in Sec. \ref{sub:Comparison of various pulse
lengths}.

\section{Results}  

\subsection{Comparison of various target states} 
\label{sub:Comparison of various target states}

\begin{figure}[t] \centering
  \includegraphics[width=7.8cm]{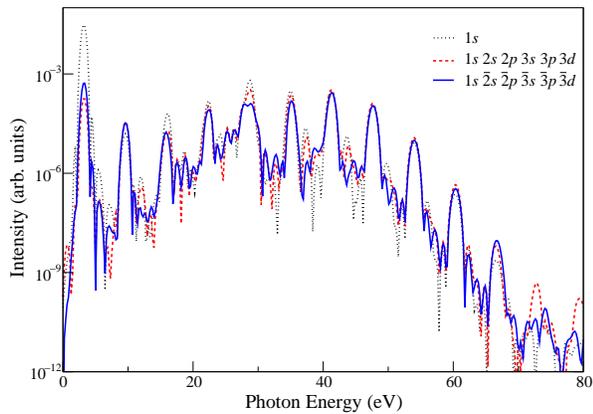} \caption{(Color online)
	 The harmonic spectrum (up to a constant of proportionality) as calculated
	 from the dipole acceleration for He in a 390 nm, $4\times
	 10^{14}$ Wcm$^{-2}$, 3-2-3 laser field, using as a model residual ion description, the
	 $1s$ state (black, dotted line), the $1s$, $2s$, $2p$, $3s$, $3p$, $3d$ states (red, dashed
	 line), and the $1s,\overline{2}s,\overline{2}p,
	 \overline{3}s,\overline{3}p,\overline{3}d$
	 pseudostates (blue, solid line). The single state model provides a
	 reasonable approximation to the more detailed descriptions beyond the first
	 harmonic where there is a large discrepancy between the spectra.
	 \label{fig:all-targ-acc-comparison}} \end{figure}

The harmonic response, as calculated from the expectation value of the dipole
acceleration, of a He target in the 1T, 6T and 6P configurations is shown in
Fig.  \ref{fig:all-targ-acc-comparison}. The spectra display the expected form-
a pronounced first harmonic peak followed by a plateau of peaks at odd multiples
of the fundamental photon energy, which decay exponentially beyond a cutoff.
The cutoff of the plateau appears at a photon energy of approximately 45 eV. The
standard formula for the cutoff energy, $I_p+3.2U_p$ \cite{cutoff_law}, where
$I_p$ is the ionization potential and $U_p$ the ponderomotive energy, is not
necessarily appropriate in this wavelength and intensity regime.  Nevertheless,
for the current parameters, it predicts a cutoff energy of 42 eV.  The observed
cutoff is therefore not inconsistent with the cutoff formula.  

We can compare the spectra to assess how the description of atomic structure
affects the calculated HG spectra. The 1T calculations are in better agreement
with the more detailed calculations at higher energies, especially in the cutoff
region between the 13th and 19th harmonics where agreement is within 30\%.  In
the low energy region, and especially in the first harmonic, the spectra differ
significantly- the first harmonic response in the 1T model is 60 times greater
than that in the 6P model. The inconsistencies in the lower harmonics 
between the 1T and the 6T and 6P models imply that the low energy harmonics in
the dipole acceleration calculation are highly sensitive to changes in the
atomic structure, and that in the higher energy cutoff region the details of the
atomic structure are not as important.  There is a factor 3 difference in the
first harmonic peak between the 6T and 6P models. As pseudostates may better
represent the changes to the ground state than true states, this difference
implies that the first harmonic is especially sensitive to the description of
the ground state. The 6P spectrum shows a double peak structure at the 9th
harmonic stage which the two true state models do not.  

\begin{figure}[t] \centering
  \includegraphics[width=7.8cm]{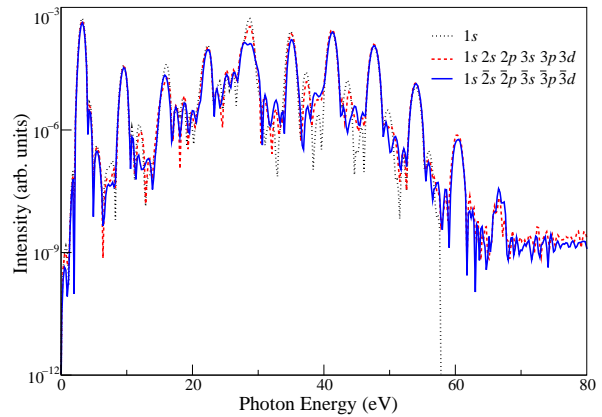} 
  \caption{(Color online) The length form harmonic spectrum of He in a
	 390 nm, $4\times 10^{14}$ Wcm$^{-2}$, 3-2-3 laser field, using as a
	 residual ion description, the 1T (black, dotted line),  6T (red, dashed line), and 6P (blue,
	 solid line) models (See Fig. \ref{fig:all-targ-acc-comparison} for details).
	 Results from the 1T model provide a reasonable approximation to results from the more detailed
	 descriptions especially in the cutoff region.  There is also good agreement for the
	 first harmonic when compared with the large discrepancy in Fig.
	 \ref{fig:all-targ-acc-comparison}. \label{fig:all-targ-len-comparison}}
  \end{figure}

As the first term in the dipole acceleration is proportional to $1/r^2$ it is
most sensitive to changes in the wavefunction at small $r$. If the description
of the atomic structure close to the nucleus is not exact, this can lead to
significant inaccuracies in the low energy region of the spectra calculated from
the dipole acceleration, especially the first harmonic peak. Figure
\ref{fig:all-targ-len-comparison} shows the same harmonic spectra as Fig.
\ref{fig:all-targ-acc-comparison} but in this case the spectra are calculated
from the dipole length operator. In this form the harmonics are far less
sensitive to the details of the atomic structure close to the nucleus as can be
seen by the excellent agreement between the three spectra at the first and third
harmonics (within 20\%). 

In fact the agreement between the spectra from the
different target states is generally better in the length and velocity forms
than in the acceleration- except for the 9th and 11th harmonics, the agreement
between the 6T and 6P dipole-length spectra is within 20\%. This further
highlights that the dipole acceleration is especially sensitive to the
description of atomic structure. The main difference between the three spectra
appears again in the 9th harmonic. This difference is very similar to the
difference seen in Fig.  {\ref{fig:all-targ-acc-comparison}} in which the dipole
acceleration was used to determine the harmonic spectrum. This indicates that
this difference originates from the different bases used, rather than the choice
of operator for the determination of the harmonic spectrum. This topic will be
discussed further in section \ref{sub:Comparison with HELIUM}.

\subsection{Comparison of dipole length, dipole velocity and dipole acceleration forms} 
\label{sub:Comparison of various gauges}

As has been addressed in the previous section, TDRM theory can calculate
harmonic spectra from the dipole length, dipole velocity or dipole acceleration
operators. We have already seen how the dipole acceleration is sensitive to the
description of the atomic structure, particularly when it comes to the low
energy region of the spectrum. 

\begin{figure}[t]
	\centering
	\includegraphics[width=7.8cm]{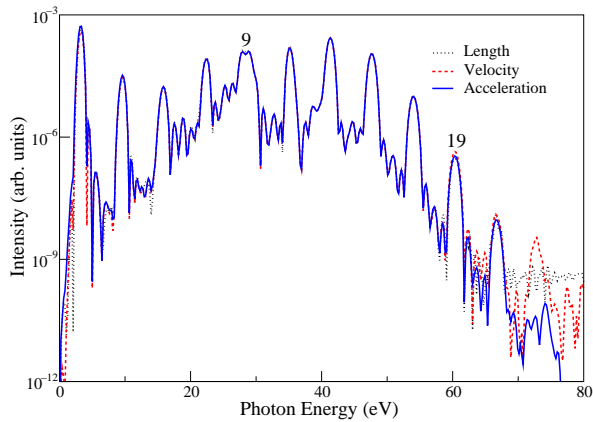}  
	\caption{(Color online) The harmonic spectrum of a pseudostate (6P) He
	  target in a 390 nm, $4\times 10^{14}$ Wcm$^{-2}$ 3-2-3 laser field, as
	  calculated from the dipole length (black, dotted line), velocity (red, dashed line) and
	  acceleration (blue, solid line). Agreement to within 20\% is found
	  between all three spectra up to the 19th harmonic peak (60 eV). The spectra
	  diverge beyond this. \label{fig:6P_gauge_comparison}} 

\end{figure}

Figure \ref{fig:6P_gauge_comparison} shows the harmonic spectrum of 6P He in a
390 nm, $4\times 10^{14}$ Wcm$^{-2}$ laser field as calculated using the dipole
length, velocity and acceleration forms of the dipole matrix elements.  The
pseudostates model gives a more accurate description of the changes in the
ground state due to the laser pulse, and hence should give a more accurate
picture of the harmonic spectrum than the true state model. In terms of the
agreement between the spectra this holds true, as the 6P model gives a
consistent agreement between the three different approaches to calculate the
harmonic spectrum where the 6T model breaks down at low harmonics.  For the 6P
model the three spectra agree within 20\% at every harmonic peak up to the 19th,
well into the cutoff region. In the 6T there is agreement within 20\% between
the dipole length and velocity spectra, but the dipole acceleration spectrum
differs by 60\%, 30\% and 40\% in the first, third and fifth harmonics
respectively.

\begin{figure}[t]
  \centering
  \includegraphics[width=7.8cm]{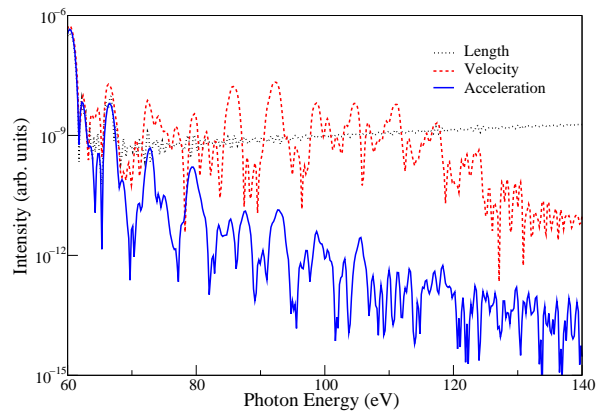}
  \caption{(Color online) The high energy harmonic spectrum of true state (6T)
	 He target in a 390 nm, $4\times 10^{14}$ Wcm$^{-2}$ 3-2-3 laser
	 field, as calculated from the dipole length (black, dotted line), velocity (red, dashed
	 line) and acceleration (blue, solid line). The leftmost harmonic
	 shown is the 19th, above which the spectra diverge.  \label{fig:6T_gauge_comparison}} 
	 
\end{figure}

Regardless of which model is used, the three spectra diverge beyond the 19th
harmonic (Fig. \ref{fig:6T_gauge_comparison}), with the dipole length spectrum
becoming noisy and the dipole acceleration spectrum displaying a few more weak
harmonics decaying into noise. The dipole velocity spectrum on the other hand
displays a second plateau of peaks not seen in the other spectra. These peaks
are not predicted classically, and their absence from the other spectra implies
that they are spurious. This implies that the length and velocity forms are
reliable, but only in an energy range up to and including the cutoff region.
This is especially important as for general multielectron targets the
acceleration form will be prohibitively sensitive to the limitations in the
description atomic structure (See Sec.  \ref{sec:calcparam}). However, by using
both the dipole length and dipole velocity operators it is possible to obtain
reliable harmonic spectra for multielectron systems using the TDRM approach.

\subsection{Comparison with HELIUM} 
\label{sub:Comparison with HELIUM}

Having demonstrated that the TDRM method is self consistent within a certain
energy range, we now seek to benchmark our results against those from a proven
alternative method.  The HELIUM method \cite{HELIUM} uses direct numerical
integration of the full-dimensional TDSE to describe a two-electron system. By
solving the TDSE directly, no significant approximations are made, and thus all
important multielectron effects are included. This makes HELIUM an excellent
code against which to benchmark TDRM.

\begin{figure}[t]
	\centering
	\includegraphics[width=7.8cm]{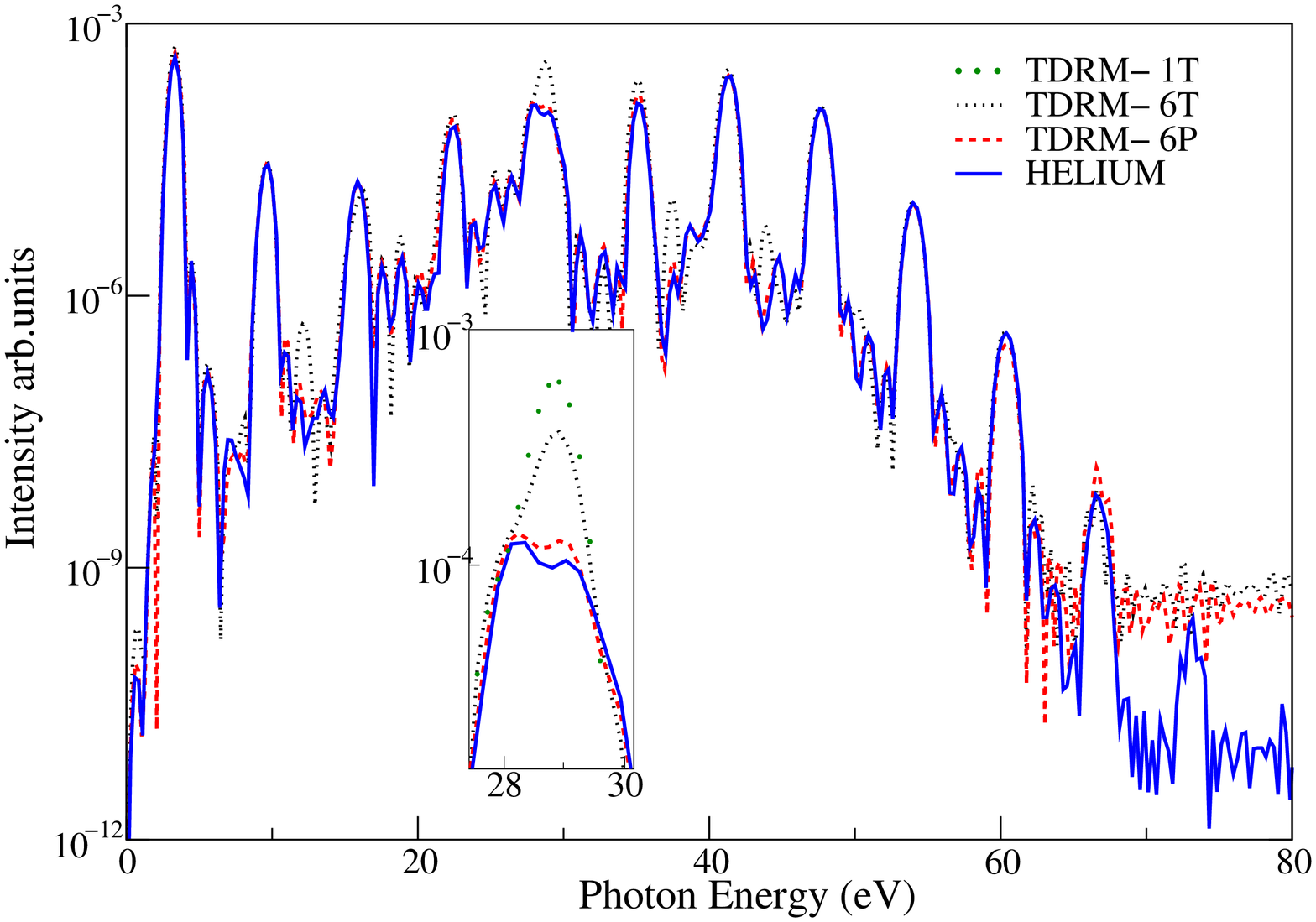}
	\caption{(Color online) The harmonic spectra as calculated from the dipole
	  length produced from the 6T (black, dotted line) and 6P (red,
	  dashed line) He models for TDRM, and from the HELIUM code (blue, solid
	  line). The 6P and 6T spectra agree with the HELIUM spectrum to within 20\%
	  and 30\% respectively up to the 21st harmonic peak (except at the 9th and
	  11th harmonics for the 6T spectrum).
	  Inset: Both the 6P and HELIUM models have a structured peak at
	  the 9th harmonic. The 1T (green circles) and 6T spectra do not.
	  \label{fig:6TP-hel-comparison}}
\end{figure}

Figure \ref{fig:6TP-hel-comparison} shows the length form harmonic spectra
produced by the 6T and 6P models of He alongside that produced by the HELIUM
code, for a target in a 390 nm, $4\times 10^{14} $ Wcm$^{-2}$, 3-2-3 laser
field. At the harmonic peaks the agreement is very
good. The 6P and HELIUM spectra agree to within 20\% up to the 21st
harmonics, while the 6T spectrum is within 30\% except at the 9th and 11th
harmonics

The inset in Fig. \ref{fig:6TP-hel-comparison} shows detail in the 9th harmonic
from the three calculations, and the TDRM 1T model. The 6P model and HELIUM
spectra show a structured peak which the 1T and 6T do not. The ponderomotive
energy in the laser field shifts the He ground state down by around 5.7 eV,
shifting the $1s3p$ bound state into the vicinity of the 9th harmonic peak. The
presence of a bound state has been shown to give rise to such structure in the
harmonic peaks \cite{brown_prl}. It is useful to notice that the 6T model may
not describe the changes to the He$^+$ ground state in the laser field as
accurately. Thus, the shift of the $1s3p$ state peak may differ and consequently
we do not observe the double peak structure in the 9th harmonic for the 6T
spectrum.  The 1T model does not account for any changes to the He$^+$ ground
state, and differences between the 1T model and the other models are thus even
larger.  Expansion of the basis set in the TDRM approach thus leads to a
harmonic spectrum which gets closer to the benchmark harmonic spectrum obtained
using the HELIUM code.

\begin{figure}[t]
	\centering
	\includegraphics[width=7.8cm]{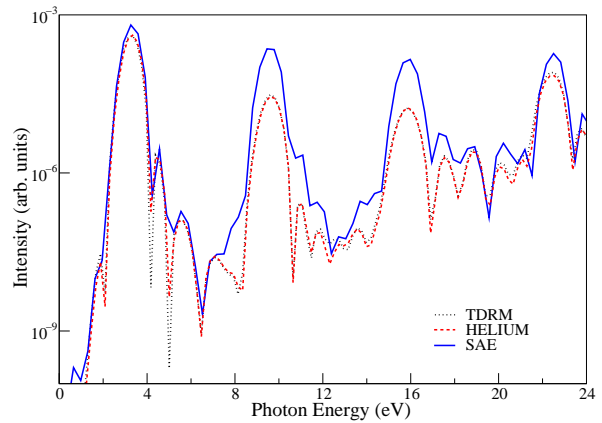}
	\caption{(Color online) The low energy region of the dipole velocity harmonic
	  spectrum produced from the 6P He model in TDRM (black, dotted line),
	  and the length form harmonic spectra from the HELIUM code (red, dashed
	  line), and its SAE derivative (blue, solid line). The TDRM and HELIUM spectra are
	  indistinguishable, but the SAE spectrum overestimates the harmonic spectrum
	  at low energies. \label{fig:td-hel-sae}}
\end{figure}

The agreement for the TDRM velocity form spectrum is even better: within 15\%
when comparing the velocity form, 6P, TDRM spectrum with the length form HELIUM
spectrum.  The excellent agreement between the spectra serves to give weight to
the results obtained from both methods. The sensitivity of the harmonic spectra
to the description of atomic structure makes it even more remarkable that the
two methods overlap, especially in the low energy region. Figure
\ref{fig:td-hel-sae} shows the low energy region of the velocity form spectrum
obtained from the 6P model TDRM code, alongside the length form spectra from
the HELIUM code and from an SAE simplification derived from the HELIUM code
\cite{helsae}. The three spectra agree well in the first harmonic, whereas the
acceleration form spectra (not shown) vary widely. This confirms that the
dipole velocity and length are significantly less sensitive to the description
of atomic structure close to the nucleus, and are probably more reliable in the
low energy, especially first harmonic, region. The gridspacing in HELIUM and
the limited basis set in TDRM impose constraints on the calculations very close
to the nucleus. These constraints make it likely that the acceleration form
spectra are less reliable in the first harmonic, which could give rise to the
discrepancy between the two methods.

In the third and fifth harmonics the SAE model markedly overestimates the
harmonic spectra obtained from both the TDRM and HELIUM models, which are in
excellent agreement with each other. This implies that the SAE model is not
sufficient to describe low energy harmonic spectra, and that the lowest energy
harmonics are significantly more sensitive to atomic structure.  We note that
in the plateau and cutoff regions the SAE spectrum is in good agreement with
the full HELIUM spectrum, lending justification to the use of the SAE
approximation for investigating the generation of higher harmonics in He.



\subsection{Comparison of various pulse lengths} 
\label{sub:Comparison of various pulse lengths}

\begin{figure}[t]
	\centering
	\includegraphics[width=7.8cm]{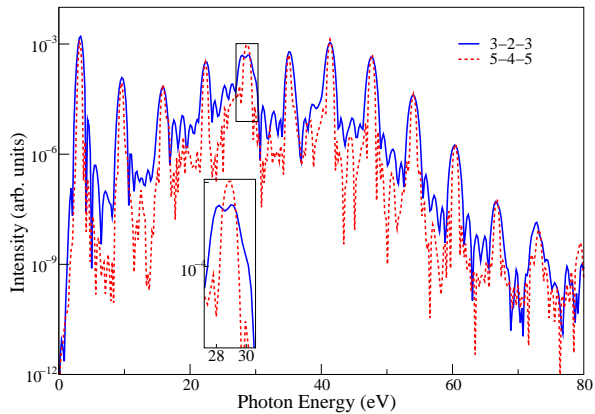}
	\caption{(Color online) The dipole velocity harmonic spectra produced from the
	  6P He model for a 3-2-3 (blue, solid line) (3 cycle, $\sin ^2$ ramp
	  on, 2 cycles peak intensity, 3 cycle, $\sin^2$ ramp off) and a 5-4-5 pulse
	  (red, dashed line).  Both pulses have a peak intensity of $4\times
	  10^{14}$ Wcm$^{-2}$ and a wavelength of 390 nm. The longer 5-4-5 pulse
	  gives rise to narrower harmonic peaks, but the peak values are still within
	  20\% of the 3-2-3 spectrum. Inset: There is a significant difference
	  between the two spectra in the 9th harmonic. \label{fig:pulse-comparison} }
\end{figure}

To probe the effect of the laser pulse profile on the harmonic spectra, as well
as the 3-2-3 (3 cycles, $\sin^2$ ramp on, 2 cycles peak intensity, 3 cycles
$\sin^2$ ramp off) profile, we ran calculations for various longer pulses,
namely 5-2-5, 3-4-3 and 5-4-5 pulses. Broadly speaking, while the spectra
themselves change (with narrowing peaks for the longer pulses), the comparisons
between the 1T, 6T and 6P  models, with the HELIUM results or between the
dipole length, velocity and acceleration forms do not change significantly.
Figure \ref{fig:pulse-comparison} shows the spectra produced by a 3-2-3 and a
5-4-5 laser pulse. The peak values are not changed significantly by the
different pulse profile, but the longer 5-4-5 pulse gives rise to narrower peaks
and greater contrast.  This gives a greater energy resolution between different
peaks. Therefore the broad 9th harmonic peak in the 3-2-3 spectrum in Fig.
\ref{fig:pulse-comparison} is further broadened by the presence of the nearby
$1s3p$ bound state, whereas the narrower peak arising from the 5-4-5 pulse is
isolated from any nearby atomic structure. 

\section{Conclusions} 
\label{sec:Conclusions}

We have extended the calculation of harmonic spectra in TDRM theory by
determining these spectra through the time-varying expectation value of the
dipole length, dipole velocity and dipole acceleration operators, and applied
the adapted codes to He irradiated by a 390 nm, $4\times10^{14}$ Wcm$^{-2}$
laser field. We have compared the spectra calculated using each form, assessed
the effect of changing the multielectron basis set used to describe the residual ion,
and benchmarked our results against those obtained from the HELIUM method. 

We have shown that for harmonic photon energies up to and including the cutoff
region the TDRM method provides results which are both self-consistent
(between dipole length, velocity and acceleration forms) and consistent with an
independent approach.  The favorable comparison between the TDRM and HELIUM
methods in the velocity and length form spectra implies that the present
approach can provide excellent results. Care must be taken in the lower
harmonics especially if using the dipole acceleration operator where the
sensitivity to inaccuracies in the description of the atomic structure can
seriously affect the reliability of the spectra obtained. 

For general multielectron systems we can perform the calculations using both the dipole length
and velocity, and compare the two spectra in order to establish bounds on
the reliability of the results. Both methods give excellent agreement for He
well into the cutoff region. The divergence of the spectra beyond this occurs at
energies which are usually outside the region of interest. 

We have also probed the advantages of the various residual ion descriptions,
which can be used within the TDRM method, finding that smaller basis sets, such
as the 1T single target state, provide an efficient way of testing the code,
and a reasonable approximation to the harmonic spectrum, but larger basis sets
give more detailed spectra, as would be expected from their better description
of the atomic structure involved. We also find that the inclusion of
pseudostates in the He$^+$ basis seems to lead to more accurate harmonic
spectra. This is particularly noticeable when compared with the highly accurate
HELIUM method.  This is largely due to the more precise way in which the
pseudostate model describes the variations in the He$^+$ ground state in
response to the laser field. 

However, the use of pseudostates for general multielectron atoms can be
problematic. By introducing non-physical thresholds into the system,
pseudo-resonances can show up in the harmonic spectrum. These inadvertant
features do not appear in the He case presented here, as the energies at which
they become important are outside the harmonic region of interest. For general
multielectron atoms, this is not necessarily the case. This does not mean that
accurate calculations are not possible for larger atoms. Pseudostates can be
used as long as care is taken- with knowledge of the position of
pseudo-thresholds, unphysical resonances can be identified and disregarded.
Secondly, although the 6T He model is not as close to the HELIUM spectrum as the
6P, it is still within 30\% at every harmonic peak except the 9th and 11th (in
the dipole length spectra).  Physical orbitals can thus also be used to improve
accuracy of harmonic spectra. The number of physical orbitals required may be
larger than if pseudo-orbitals are used, but this is not a fundamental problem:
it affects only the scale of the calculations. With careful analysis of,
and comparison between, pure physical orbital and pseudostate models we can
reliably assess the accuracy of harmonic spectra for general multielectron
systems using TDRM theory.

Furthermore, even models which use only physical orbitals already offer significant gains
over SAE models. A simple example of this is HG in Ar$^+$. Harmonics produced by
Ar$^+$ ions have been suggested to be the source of the highest harmonics
observed from a neutral Ar target \cite{argon+_gibson,argon+_zepf}. The presence
of three low-lying, $3s^23p^4$ Ar$^2+$ thresholds can have a significant
effect on the harmonic spectrum, and hence interactions between channels
associated with these thresholds must be accounted for. 
These interactions are neglected in an SAE calculation, but would be accounted for in
a TDRM calculation involving purely physical orbitals. 

We find that the reliability of the results is not significantly affected by the
particular laser pulse profile used. We compared results for four different
laser pulse profiles, finding that while the harmonic spectra differed between
cases, the changes were consistent between the various target state models, and
with the HELIUM code results. In cases where atomic structure gives rise to
structure in the harmonic spectrum the laser pulse length may affect the way in which
this is observed. The greater energy resolution afforded by longer pulses can
isolate the separate effects of atomic structure.

The results presented are also consistent with those from  various peak
intensities.  We calculated harmonic spectra for intensities between $1\times
10^{14}$ Wcm$ ^{-2} $ and $4\times 10^{14}$ Wcm$ ^{-2} $, finding
that the results are largely consistent.  At lower intensities the plateau
region is severely truncated and so it is difficult to compare between the
various spectra, but the agreement is still evident in the cutoff region. 

The TDRM method has been rigorously tested up to intensities of 4 $\times 10
^{14}$ Wcm$^{-2}$ and at wavelengths up to 390 nm, but requires a
significant amount of development to extend beyond these limits. It will be
interesting to compare these findings with those that will be determined using
the new RMT ($R$-matrix with time) codes \cite{RMT,RMT2}  which may be better
suited to address higher intensities and longer wavelengths. While the TDRM
method has been proven to provide interesting insight into the multielectron
nature of HG, it has thus far only been implemented for general multielectron
atoms using the dipole length operator \cite{brown_prl}. The next stage will be
to apply TDRM at high intensities to systems other than He.
While the dipole acceleration is too sensitive to the description of atomic
structure to accurately describe such atoms, the length and velocity forms are
stable enough to provide good results for general targets. 


\section{Acknowledgements} 
\label{sec:Acknowledgements} 

ACB and DJR acknowledge support from the Department of Employment and Learning
NI under the programme for government. HWH is supported by the EPSRC under
grant reference number EP/G055416/1. The authors would like to thank Prof K T
Taylor and Dr J S Parker for valuable discussions and assistance with the
HELIUM code calculations.


\bibliography{mybib}

\end{document}